\documentclass[iop,apjl,revtex4]{emulateapj}

\shorttitle{Spin alignment in filaments and sheets}
\shortauthors{Tempel \& Libeskind}

\begin{document}

\title{Galaxy spin alignment in filaments and sheets: observational evidence}


\author{Elmo Tempel}
\affil{Tartu Observatory, Observatooriumi 1, 61602 T\~oravere, Estonia}
\affil{National Institute of Chemical Physics and Biophysics, R\"avala pst 10, Tallinn 10143, Estonia}
\email{elmo@to.ee}

\and

\author{Noam I. Libeskind}
\affil{Leibniz-Institut f\"ur Astrophysik Potsdam, An der Sternwarte 16, D-14482 Potsdam, Germany}
\email{nlibeskind@aip.de}


\begin{abstract}
	Galaxies properties are known to be affected by their environment. One important question is how their angular momentum reflects the surrounding cosmic web. We use the SDSS to investigate the spin axes of spiral and elliptical galaxies relative to their surrounding filament/sheet orientations. To detect filaments a marked point process with interactions (the ``Bisous model'') is used. Sheets are found by detecting ``flattened'' filaments. The minor axes of ellipticals are found to be preferentially perpendicular to hosting filaments. A weak correlation is found with sheets. These findings are consistent with the notion that elliptical galaxies formed via mergers which predominantly occurred along the filaments. The spin axis of spiral galaxies is found to align with the host filament, with no correlation between spiral spin and sheet normal. When examined as a function of distance from the filament axis, a much stronger correlation is found in outer parts, suggesting that the alignment is driven by the laminar infall of gas from sheets to filaments. When compared with numerical simulations our results suggest that the connection between DM halo and galaxy spin is not straightforward. Our results provide an important input to the understanding of how galaxies acquire their angular momentum.
\end{abstract}


%=============================================================================
\section{Introduction}

Large spectroscopic surveys, such as the SDSS, allow astronomers to correlate, with large statistical certainty, properties of galaxies with the environment they are embedded in. This is key to understanding how galaxies acquired their angular momentum, since the most successful analytical theory to describe this, asserts that torques imparted on collapsing proto-halos from the tidal field generates spin \citep{Hoyle:94,Peebles:69,Efstathiou:79,White:84,Porciani:02}. The theory of tidal torques \citep[see][for a review]{Schafer:09} thus implies correlations between the tidal field and galaxy spin.

Recent studies suggest that the formation of dark matter (DM) halos have two-phase angular momentum acquisition \citep{Codis:12,Aragon-Calvo:13}. Most of the low-mass halos are formed through the winding of flows embedded in misaligned walls; more massive halos are the product of mergers along such filaments. This mechanism supports the disc-forming paradigm presented by \citet{Pichon:11}.

The large-scale matter distribution in the Universe represents a complex hierarchical network of structure
often termed voids, filaments, sheets and knots, forming the so-called cosmic web
\citep{Joeveer:78,Bond:96}. Most of the galaxies and DM halos are assumed to be living in filaments that connect clusters of galaxies \citep{Pimbblet:04,AragonCalvo:10}. Since filaments are located in intersection of walls/sheets, it is important to test how the large-scale environment around galaxies is related to their angular momentum.

Using $N$-body simulations, there is a consensus that high-mass halos are preferentially oriented perpendicular to host filaments, while low-mass halos have parallel alignment \citep{Navarro:04,Aragon-Calvo:07,Brunino:07,Hahn:10,Codis:12,Libeskind:12,Aragon-Calvo:13,Trowland:13}, although \cite{Libeskind:13} suggest that the mass at which this transition occurs depends on the local density.

From an observational point of view, the picture is not as clear due to systematics and limited statistics. Several studies have tackled this problem but give somewhat contradictory results \citep{Trujillo:06,Lee:07,Paz:08,Jones:10,Cervantes-Sodi:10,Andrae:11,Varela:12}. A recent study by \citet{Tempel:13} showed a link between low-mass DM halos and spiral galaxies and high-mass halos and elliptical galaxies. Namely, the spin axes of bright spiral galaxies have a weak tendency to be aligned parallel to filaments. For elliptical galaxies, the short axes are aligned preferentially perpendicular to the host filaments.

In the present study we extend that work, by studying the correlation of galaxy orientation with respect to filaments and cosmic sheets. Since matter and gas falls into filaments from sheets, such a study will allow for an understanding of how the flow of gas/matter affects the spins of galaxies.

\newpage
%=============================================================================
\section{Observational data}
\label{sect:data}

Our study is based on the SDSS~DR8 \citep{Aihara:11}. We use the main contiguous area of the survey and the spectroscopic galaxy sample as compiled in \citet{Tempel:12}.

Due to the peculiar velocities of galaxies compact structures in redshift-space are elongated along the line-of-sight \citep{Jackson:72}. To correct for this effect and find a filamentary network, redshift-space distortions of groups are suppressed using a FOF algorithm \citep[see][]{Tago:08,Tago:10}. Groups are spherized using their rms angular sizes and their rms radial velocities as described in \citet{Liivamagi:12}.
Other redshift-space distortions are ignored when detecting filaments, however, they are implicitly taken account when calculating the randomized correlation signal (see Sect.~\ref{sect:signal}).

To estimate the spin axes of galaxies, the photometry of galaxies is modeled as described in \citet{Tempel:13}. From the galaxy modeling, the inclination and position angles for each galaxy is obtained. The latter is uniquely defined, but the inclination angle is model dependent. \citet{Tempel:13} used the apparent ellipticity of galaxies to estimate the inclination angles and showed that the alignment signals are independent of the used method.

The alignment of spiral and elliptical galaxies are examined separately. The morphological classification as described in \citet{Tempel:11} is used wherein galaxies are divided into three classes: spirals, ellipticals, and those with uncertain classification. \citet{Tempel:12} compared this classification with the classification by \citet{Huertas-Company:11} and showed that they are in very good agreement. For the current study, the samples include only galaxies that have the same classification in both papers, hence, our classification is rather conservative.

As a final step, galaxies which are close to the borders of the survey are eliminated since reliable filaments cannot be detected there. The limiting distance from the survey border is set to $3\,h^{-1}\mathrm{Mpc}$: a spatial sample mask is used as defined in \citet{Martinez:09}.

As in a previous paper \citep{Tempel:13}, only the correlation signal for bright galaxies ($M_r<-20.5\,\mathrm{mag}$) is measured. Fainter galaxies are neglected, because the estimation of their spin axes is uncertain. This is in part also motivated by \citet{Lee:04} who showed that the intrinsic correlation is stronger for brighter galaxies. Note also that only galaxies that are in filaments (namely, that are closer than $0.5\,h^{-1}\mathrm{Mpc}$ to the filament axis) are used. The total number of bright elliptical/spiral galaxies found in filaments and used here is 7433/5283.

%=============================================================================
\section{Methods}
\label{sect:methods}

%----------------------------------------------
\subsection{Filament finder}

Filaments are traced by applying an object point process with interactions (the Bisous process) to the distribution of galaxies. Accordingly, random segments (thin cylinders) are placed on the galaxy distribution. If two segments are aligned and linked the probability of finding a filament is increased. The morphological and
quantitative characteristics of these complex geometrical objects can be obtained
by following a straightforward procedure: constructing a model, sampling the
probability density describing the model, and, finally, applying the methods of
statistical inference. We use the Metropolis-Hastings algorithm together with simulated annealing to sample the model probabilities.  After a large number of repetitions of the process, a network of filaments emerges, each filament being labelled with its coordinates, direction, and statistical significance.

A detailed description of these methods is given in \citet{Stoica:07,Stoica:10}. The exact realization of the Bisous process as used in the present study is described in \citet{Tempel:13a}. In practice, after fixing the approximate scale of the filaments, the method returns a filament detection probability and filament orientation fields. Every detected structure in this model is a filament (by definition), and the detection probability describes the relative strength of the structure. Based on the detection probability field and the orientation field, filament axes are revealed. 

This method requires the setting of characteristic scale~-- corresponding to the filament radius~-- which is set to $0.5\,h^{-1}\mathrm{Mpc}$. It is the minimum scale on which reliable measurements can be made and it roughly corresponding to the size of a galaxy group/cluster. We expect that the searched alignment should be present on that scale. Naturally, the nature of filaments is hierarchical \citep{Shen:06,Aragon-Calvo:13} and the chosen scale can be arbitrary.

%----------------------------------------------
\subsection{Estimating the sheet orientations}

Our filament finder is not able to detect cosmic sheets directly. However, since galaxy filaments are embedded in sheets, their orientation in filaments can be estimated. In order to do so the probabilistic nature of the filament finder is employed. In practice, after the filament detection probability field is found the cross section of the probability field (at filament axes) may be examined: at the location of filament axes, the probability maps have a maximum. If a filament is located in sheet, the probability maximum is elongated in the direction of galactic sheet. We note that the orientation of sheets defined this way can be only used statistically.

The direction along which the filament maxima are elongated is used to define the sheet orientation. Following the convention from the velocity shear tensor analysis \citep{Hoffman:12}, let the filament axis vector be $\mathbf{e}_{3}$ and the sheet normal vector be $\mathbf{e}_{1}$; $\mathbf{e}_{2}$ is perpendicular to these two vectors.

\citet{Tempel:13b} showed that the orientation of the velocity shear vector $\hat{e}_3$ and filaments as defined by the Bisous model are strongly aligned, however, many detected filaments are located in sheets, based on velocity field classification. This is because, the present method detects flattened filaments that are classified as sheets based on velocity field \citep[see visualization in][]{Tempel:13b}. Moreover, the directions of $\mathbf{e}_{1},\mathbf{e}_{2}$ for Bisous filaments are also strongly aligned with $\hat{e}_1,\hat{e}_2$ of the velocity shear vector (in~prep).

%----------------------------------------------
\subsection{Measuring the alignment signal}
\label{sect:signal}

To estimate the orientation of galaxies relative to filaments/sheets, the probability distribution function $P(|\cos{\theta}|)$ is measured, where $\theta$ is the angle between the galaxy rotation axis and the studied vector ($\mathbf{e}_{1},\mathbf{e}_{2},\mathbf{e}_{3}$). The quantity $\cos{\theta}$ is obtained as a scalar product between the two unit vectors: $\cos{\theta}=1$ implies that the galaxy spin is parallel to $\mathbf{e}_{i}$, while $\cos{\theta}=0$ indicates it is perpendicular.

The probability distribution function should be compared with the null-hypothesis of random mutual orientation of galaxies and vectors. Due to selection effects, this is not simply a uniform distribution; neither the inclination angles of galaxies nor the distribution of filament axes (with respect to the line-of-sight) have random orientations \citep[see][]{Tempel:13}. A Monte-Carlo approximation is used to estimate the distribution of $|\cos(\theta)|$ for the case where there are no intrinsic correlations, and to find the confidence intervals for this estimate. This approach takes simultaneously into account the biases in filament detection (redshift-space distortions) and estimation of galaxy spins.

In order to do so, 10000 randomized samples are generated in which the orientations (inclination and position angles) of galaxies are kept fixed, but galaxy locations are randomly changed between filament points. This gives the true random orientation angle between the galaxy spin and filament axis. In principle, the randomized distribution depends how the filament points are chosen: based on filament axes, location of galaxies etc. However, for the current dataset it turns out to be insensitive to that. Using randomized samples the median of the null-hypothesis of a random alignment is calculated together with its 95\% confidence limits.

The galaxy spin vector is not uniquely defined since we do not know which side of the galaxy is closer to us. In order to handle this both spin vectors of a given galaxy are used. \citet{Varela:12} also tested this approach with several Monte-Carlo simulations and showed that the procedure recovers correctly the probability distribution function.

%=============================================================================
\section{Results}

%=============================================================================
\subsection{Elliptical galaxies}

\begin{figure} 
	\centering
	\includegraphics[width=84mm]{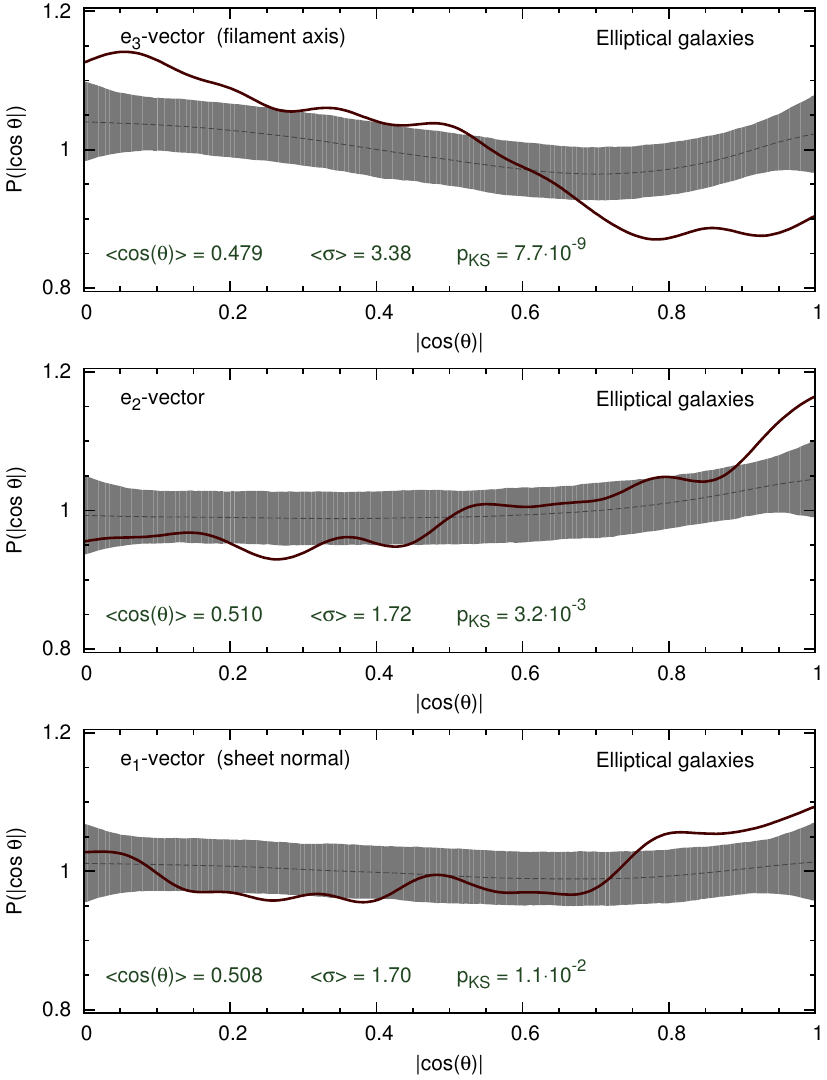}
	\caption{The orientation probability distribution between the short axes of elliptical galaxies and the filament/sheet axes. \emph{Upper panel} shows the distribution for vectors parallel to filaments; \emph{middle panel} shows the distribution for vectors perpendicular to filament but parallel to the sheet; \emph{lower panel} shows the distribution for vectors perpendicular to the sheet where filament is located. The black line and the grey filled region show the null-hypothesis together with its 95\% confidence limit. The solid red line shows the measured alignment signal.}
	\label{fig:corell}
\end{figure}

Figure~\ref{fig:corell} shows the probability distribution $P(|\cos\theta|)$ for the angle $\theta$ between the short axes of elliptical galaxies and the orientation vectors of filaments/sheets. The probability distribution is calculated for three principal vectors: $\mathbf{e}_{3}$, the filament axis; $\mathbf{e}_{1}$ the normal to the sheet where the filament is located and $\mathbf{e}_{2}$~-- a vector perpendicular to these two. In each panel of Fig.~\ref{fig:corell} we also show the average $\langle\cos(\theta)\rangle$, the average deviation from uniform distribution $\langle\sigma\rangle$ (assuming a Gaussian distribution where 95\% confidence limit corresponds to $\pm 2\sigma$) and the Kolmogorov-Smirnov (KS) test probability $p_\mathrm{KS}$ that the sample is drawn from a randomized distribution.

The alignment between filament axes and the short axes of elliptical galaxies is preferentially perpendicular as found previously \citep{Tempel:13}. Note however, that the filament finding algorithm is different~-- \cite{Tempel:13} used a locally defined morphological filtering, while here the object point process and global optimization is used. This shows that the result we obtained are rather robust and it does not depend on the filament finding algorithm (for fixed filament scale).

Moreover, estimating the short axes of elliptical galaxies is tricky since early type galaxies are triaxial ellipsoids seen in projection. Due to the degeneracy between the intrinsic oblateness of the galaxy and the inclination angle, it is nearly impossible to properly estimate a spin axis. The visible short axis of elliptical galaxies however, is easily observed, while the inclination angle is largely undefined. \citet{Tempel:13} showed that the correlation signal arises mostly from position angle of galaxies and not from inclination angle. This implies that the true alignment signal is even stronger than what we are able to measure.

The middle and lower panel in Fig.~\ref{fig:corell} show the alignment signal between the short axes of elliptical galaxies and the $\mathbf{e}_{2}$- and $\mathbf{e}_{1}$-vector, respectively. The correlation is practically the same for these two vectors. It shows that the short axes of elliptical galaxies are preferentially perpendicular to filaments and the sheet orientation is not important. 

Assuming that the short axis of an elliptical galaxy is aligned with both its spin axis and the spin of the parent DM halo \citep[however, there might be offset up to $30\degr$, see e.g.][]{Hahn:10}, our findings allow us to comment on the formation mechanism of elliptical galaxies. It is known that elliptical galaxies formed predominantly through major mergers \citep[e.g.][]{Sales:12,Wilman:13}. In mergers, the rotation axis of the resulting galaxy tends to be perpendicular to the merger direction. Our results are consistent with a picture wherein galaxies are fed with mergers that occur along the filament within which they are embedded. A similar mechanism has been proposed for the formation of high-mass DM halos \citep{Codis:12}.

%=============================================================================
\subsection{Spiral galaxies}

\begin{figure} 
	\centering
	\includegraphics[width=84mm]{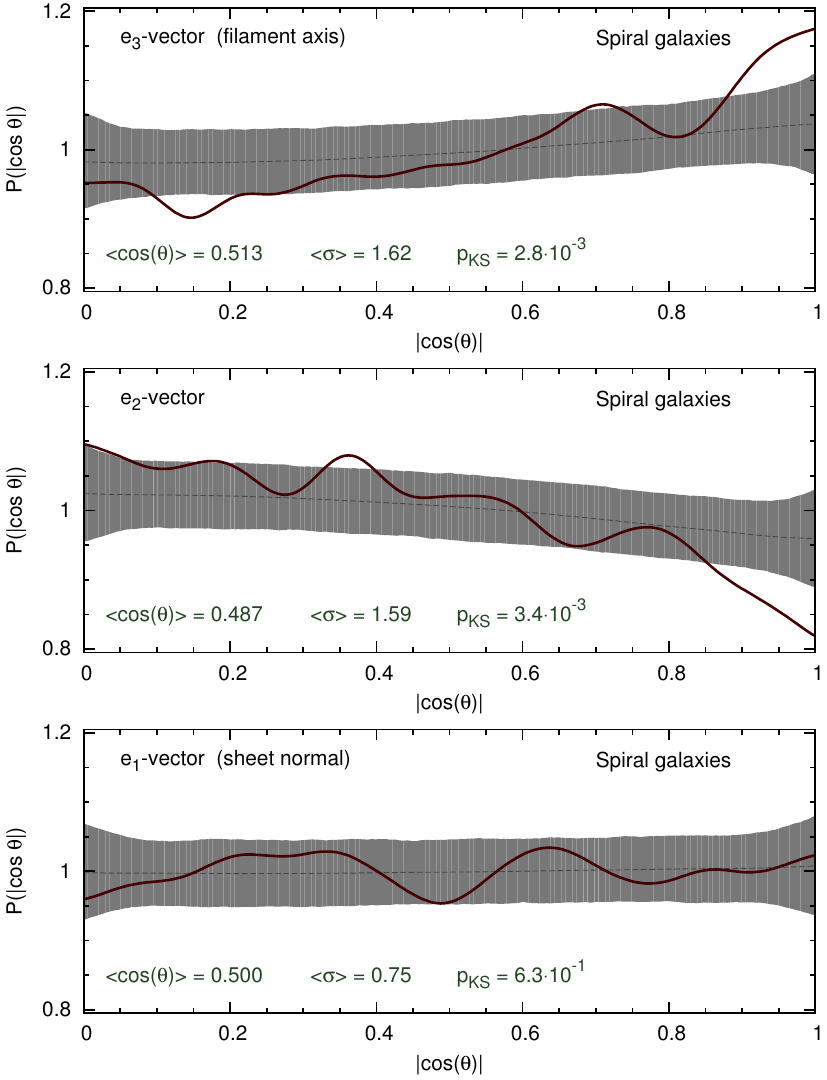} 
	\caption{The orientation probability distribution between the spin axes of spiral galaxies and the filament/sheet orientation vectors. The panels and lines are the same as in Fig.~\ref{fig:corell}.}
	\label{fig:corspi}
\end{figure}

Figure~\ref{fig:corspi} shows the correlation for spiral galaxies. The lines and designations are the same as in Fig.~\ref{fig:corell}. Figure~\ref{fig:corspi} shows that the spin axes of spiral galaxies tend to align with filaments (\emph{upper panel}), which is consistent with previous results \citep{Tempel:13}. The \emph{middle panel} of Fig.~\ref{fig:corspi}, indicates that the spin axes of spirals are preferentially perpendicular to the $\mathbf{e}_{2}$-vector. The amount of correlation is statistically the same as for the $\mathbf{e}_{3}$-vector. The \emph{lower panel} of Fig.~\ref{fig:corspi} shows that there is no statistically significant correlation between the $\mathbf{e}_{1}$-vector (sheet normal) and the spin axes of spiral galaxies. This implies that the formation of spiral galaxies is driven by the plane of the sheet along which most of the matter/gas falls in to the filaments.

\begin{figure} 
	\centering
	\includegraphics[width=86mm]{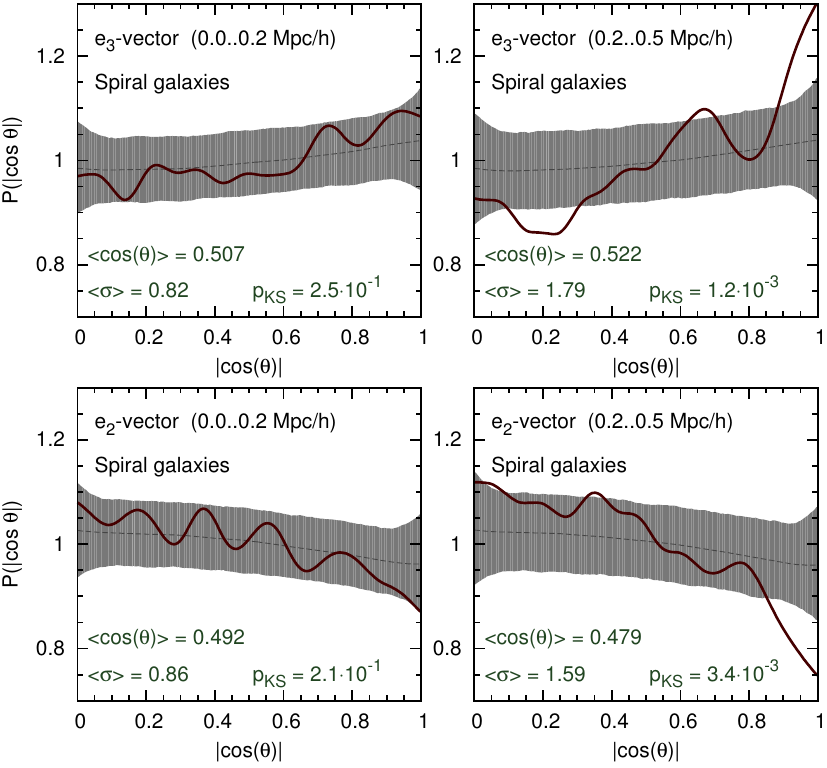}
	\caption{The orientation probability distribution between the spin axes of spiral galaxies and the filament/sheet axes. Left (right) column shows the alignment signal for galaxies that are close to (slightly away from) the filament axis. Upper/lower panels show the correlation for $\mathbf{e}_3$-/$\mathbf{e}_2$-vector.}
	\label{fig:corspidist}
\end{figure}

Figure~\ref{fig:corspidist} shows the correlation between the spin of spiral galaxies and $\mathbf{e}_{2},\mathbf{e}_{3}$ as a function of distance to the filament axis. Correlations are considerably stronger (based on KS-test probabilities) for galaxies that are slightly further away (in the range 0.2--0.5\,$h^{-1}$Mpc) than those that are closer (0--0.2\,$h^{-1}$Mpc) to the filament axis, which are consistent with random. This implies that the correlations seen above are actually driven by those galaxies slightly further way from the main filament axis. This is consistent with the idea that the origin of the alignment of angular momentum is related to the regions outside filaments, namely sheets, where most of the gas is falling in from. Along filament axes more chaotic motions dominate. \citet{Codis:12} also shows that the correlation between the rotation axes of DM halos and filaments is stronger in outer parts of filaments, supporting our findings.
 
%=============================================================================
\section{Discussion and conclusions}

We have examined the alignment of spiral/elliptical galaxies with respect to the large-scale cosmic filamentary network. The correlation signal is calculated only for bright galaxies that are located in filaments, where we also estimate the sheet orientation. The alignment between galaxy spins and the axis of filaments/sheets is characterized by the shape of the probability distribution of $\cos\theta$, where $\theta$ is the angle between the two vectors.

A significant correlation between the short axes of elliptical galaxies and filament axes is found (the KS-test $p$-value is $7.7\cdot 10^{-9}$); these galaxies tend to be spinning perpendicular to the filament axis. For bright spiral galaxies on the other hand the opposite is found: they tend to be aligned with the host filament axis. Both these results confirm earlier findings which employed different filament finding algorithms \citep{Tempel:13}.

In this study, no alignment between the spin axes of spiral galaxies in filaments and the $\mathbf{e}_{1}$-vector (sheet normal) is found.

A basic interpretation of filament formation suggests that as a matter flows towards filaments, it wraps its up, thus aligning the filament axis with its angular momentum \citep[as well as the vorticity of the filamentary matter, see][]{Libeskind:13b}. Spiral galaxies which condense out of filaments should thus preserve the perpendicular alignment between their spin and the direction of matter infall. If gas infall from sheets to filaments is laminar, it gives the parallel alignment between the spin axes of spiral galaxies and orientation of filaments. Assuming that more matter flows along $\mathbf{e}_{2}$ then $\mathbf{e}_{1}$, our results confirm this picture observationally\footnote{Note that $\mathbf{e}_{1}$ is the axis along which matter is compressed the most, not the axis along which most matter flows.}. A similar scenario is observed in low-mass DM halos that form through quiet accretion of ambient matter \citep{Codis:12}.

That no correlation is found between the $\mathbf{e}_{1}$-vector and spin axes of spirals is surprising given that such correlations are measured in $N$-body simulations \citep{Aragon-Calvo:07,Libeskind:13}. There are a number of possible explanations for this null-result. One may be that since the correlations published in numerical studies are weak, the present work may simply be too limited by observational constraints to pick up the signal. 

Another explanation is that the spin of low-mass galaxies is more poorly aligned with the dark halo's spin than high-mass galaxies \citep{Hahn:10}. This may be because of stronger susceptibility to feedback \citep{Sales:12} or to spin flips \citep{Bett:12}.  The angular momentum of spiral galaxies may be somehow redistributed from within the plane of the sheet where it was ``born''. The angular momentum that is at first aligned with the $\mathbf{e}_{2}$ and perpendicular to the $\mathbf{e}_{1}$-vector may be torqued towards the $\mathbf{e}_{3}$-vector. Although we do not have the full quantitative explanation for this effect, it may be related with the fact that gas has a laminar infall from sheets to filaments. This is also supported by the fact that the alignment of spiral galaxy spin with filament axis is dependent on the distance from the filament axis~-- a stronger correlation is found in the outer parts than in the inner parts.

For elliptical galaxies, the spin correlation is the same for $\mathbf{e}_{1}$- and $\mathbf{e}_{2}$-vectors, indicating that mergers occur mostly along the filament axis and the sheet orientation does not have any impact on preferred merger direction.

Our analysis shows that the scale 0.5\,Mpc is important for tidal torqueing process. It requires a special study to find out what is the dominant scale that affects the galaxy formation.

Comparing with $N$-body simulations \citep{Aragon-Calvo:07,Libeskind:12,Codis:12}, where the angular momentum of DM halos is examined with respect to the cosmic web, our results are slightly different. Many of these studies show a stronger alignment with respect to the $\mathbf{e}_{1}$-vector. Observationally, the orientation of spiral galaxy spin axis with respect to the $\mathbf{e}_{1}$-vector is statistically consistent with random. The alignment of the spin axes of spiral galaxies is more closely related with the gas physics and less a result of the DM halo's angular momentum. This is supported by the results of \citet{Lee:11} which  suggest that the correlation between galaxies and the large-scale structure reflects the initial alignment caused by tidal torqing more than the DM halos. One way to test this, is to look the DM halos in higher redshift to see how the spin of DM halos is correlated with respect to the cosmic structures.

To resolve the discrepancy between simulations and observations, we plan to use the local Universe constrained simulation \citep{Hess:13} to study the correlation simultaneously for galaxies and DM halos.

\acknowledgments

ET acknowledge the ESF grants MJD272, SF0060067s08, and the European Regional Development Fund. NIL acknowledges a grant from the \textit{Deutsche Forschungs Gemeinschaft}. This work was carried out in the High Performance Computing Centre of University of Tartu. We thank the SDSS-III Team for the publicly available data releases (\url{http://www.sdss3.org/}).

%\bibliographystyle{apj}
%\bibliography{mybib}{}

%\clearpage

\end{document}